\pdfoutput=1

\documentclass[
    ,final            
  ]
  {aipproc}

\layoutstyle{6x9}

\usepackage{natbib}

\def\cgs{ ${\rm erg~cm}^{-2}~{\rm s}^{-1}$ }    
\def\es{ ${\rm erg~\rm s}^{-1}$ }    
\def\cm{ ${\rm cm~\rm}^{-2}$ }    
\def\gtrsim{\mathrel{\hbox{\rlap{\hbox{\lower4pt\hbox{$\sim$}}}\hbox{$>$}}}} 
 
\begin{document}

\title{Heavily obscured AGN with SIMBOL-X}

\classification{95.85.Nv; 98.54.Cm; 98.70.Qy; 98.80.Es}
\keywords      {X-ray; Active Galaxies; X-ray sources; Observational Cosmology}

\author{R. Della Ceca}{
  address={INAF-Osservatorio Astronomico di Brera, Milano}
}

\author{A. Caccianiga}{
  address={INAF-Osservatorio Astronomico di Brera, Milano}
}

\author{P. Severgnini}{
  address={INAF-Osservatorio Astronomico di Brera, Milano}
}

\begin{abstract} 
By comparing an optically selected sample of narrow lines AGN with an X-ray selected sample of AGN we have recently derived an estimate of the intrinsic 
(i.e. before absorption) 2-10 keV luminosity function (XLF) of Compton Thick AGNs. We will use this XLF to derive the number of Compton Thick AGN that will be found in the SIMBOL-X survey(s). 
\end{abstract}

\maketitle


\section{Introduction}

The discovery of quiescent Super-Massive Black-Holes (SMBH) in the nuclei of non-active nearby galaxies with prominent bulges (\cite{kormendy1995},
\cite{magorrian1998}), along with the discovery of scaling relations between the central BH mass and galaxy properties (e.g. bulge luminosity/mass and velocity dispersion, \cite{gebhardt2000}, \cite{ferrarese2000}) strongly suggest a tight link between BH growth  and galaxy evolution (commonly referred to as "co-evolution") and a physical mechanism linking accretion and ejection of material occurring in the central (sub-parsec scale) galaxy nucleus to the rest of the galaxy. This implies that AGN are not only sources where high energy physical processes take place but are leading actors in the formation and evolution of galaxies and, in general, of cosmic structures in the Universe. 

However the largest fraction of the AGN population is obscured by a large amount of cold matter around the Active Nuclei that does not permit a direct view to the central energy source. 
For absorbed AGN having column densities ($N_H$) up to few times $10^{23}$\cm the current missions at E<10 keV (XMM-Newton and Chandra) are producing a wealth of useful data that can be used to evaluate their physical and cosmological properties (e.g. see \cite{dellaceca2008} and \cite{ebrero2009} and references therein). 
On the contrary, the situation is almost completely unconstrained for the sources having $N_H$ greater than $10^{24}$\cm (the so called Compton Thick AGN, where the matter is optically thick to Compton scattering).  
According to the synthesis modeling of the cosmic X-ray background (XRB) (\cite{treister2005}, \cite{gilli2007}), these objects should represent $\sim$50\% of the total absorbed AGN population, although recent revisions (\cite{treister2009}) seem to 
suggest a smaller contribution.

In \cite{dellaceca2008b} we have reviewed the Compton Thick AGN studied so far (using the current and past X-ray missions, i.e. ASCA, BeppoSAX, XMM-Newton, Chandra, INTEGRAL, SWIFT and SUZAKU) and  discussed the X-ray spectral improvements offered by the Simbol-X mission (see also the contributions of W. Beckmann, F. Panessa and F. Fiore at this meeting). We discuss here their expected cosmological properties 
(LogN-LogS and {\it z} distribution). ($H_o$,$\Omega_M$,$\Omega_{\Lambda}$)=(65,0.3,0.7) are used in this paper. 
Finally we are considering here only Compton Thick
AGN with $N_H$ between $10^{24}$ and $10^{25}$ cm$^{-2}$.

\section{The XLF of Compton-Thick AGN}

In \cite{dellaceca2008} by comparing the results obtained using an hard (4.5-7.5 keV) X-ray selected sample of AGN with those derived by using  an optically selected sample of narrow lines AGN (\cite{simpson2005}), we have evaluated, in an indirect way,  the intrinsic (i.e. before the absorption) XLF of Compton Thick AGN (at z=0) in the 2-10 keV energy range. This XLF (thick solid line in figure 1, left panel) is described  by a smoothly connected two power-law function, ${d \Phi(L_x,z=0) \over d Log L_x}  = {A [({L_x \over L_{\star}})^{\gamma_1} + ({L_x \over L_{\star}})^{\gamma_2}]^{-1}}$, with  $\gamma_1=1.55$, $\gamma_2=2.61$, Log $L_{\star} = 44$ and A = $4.79\times 10^{-7}$ $h^3_{65}$ $Mpc^{-3}$. 
A comparison with the results from other projects is also reported in figure 1 (left panel). As can be clearly noted 
the different estimates of the space density of Compton Thick AGN are consistent (roughly within a factor 2, see the two dashed lines) with the proposed XLF of Compton Thick AGN, with a possible hint of a XLF flattening below few times $10^{42}$ ${\rm erg~\rm s}^{-1}$;
we emphasize that this agreement is remarkably good if we consider that the different estimates have been derived using very different selection criteria and methods.
Finally the consideration that the total number of Compton Thick AGN cannot be increased arbitrarily, without violating the limits imposed by the local black mass density derived from dynamical studies of local galaxy bulges\cite{marconi2004}, allows us to put some constraints on the space density of optically/infrared-{\it elusive} Compton Thick AGN; this limit is violated if the total density of Compton Thick AGN (including also the optically-elusive ones) is above the upper dashed envelope reported in figure 1 (left panel, see \cite{dellaceca2008} for details). 

\begin{figure}
   \includegraphics[width=5.5cm]{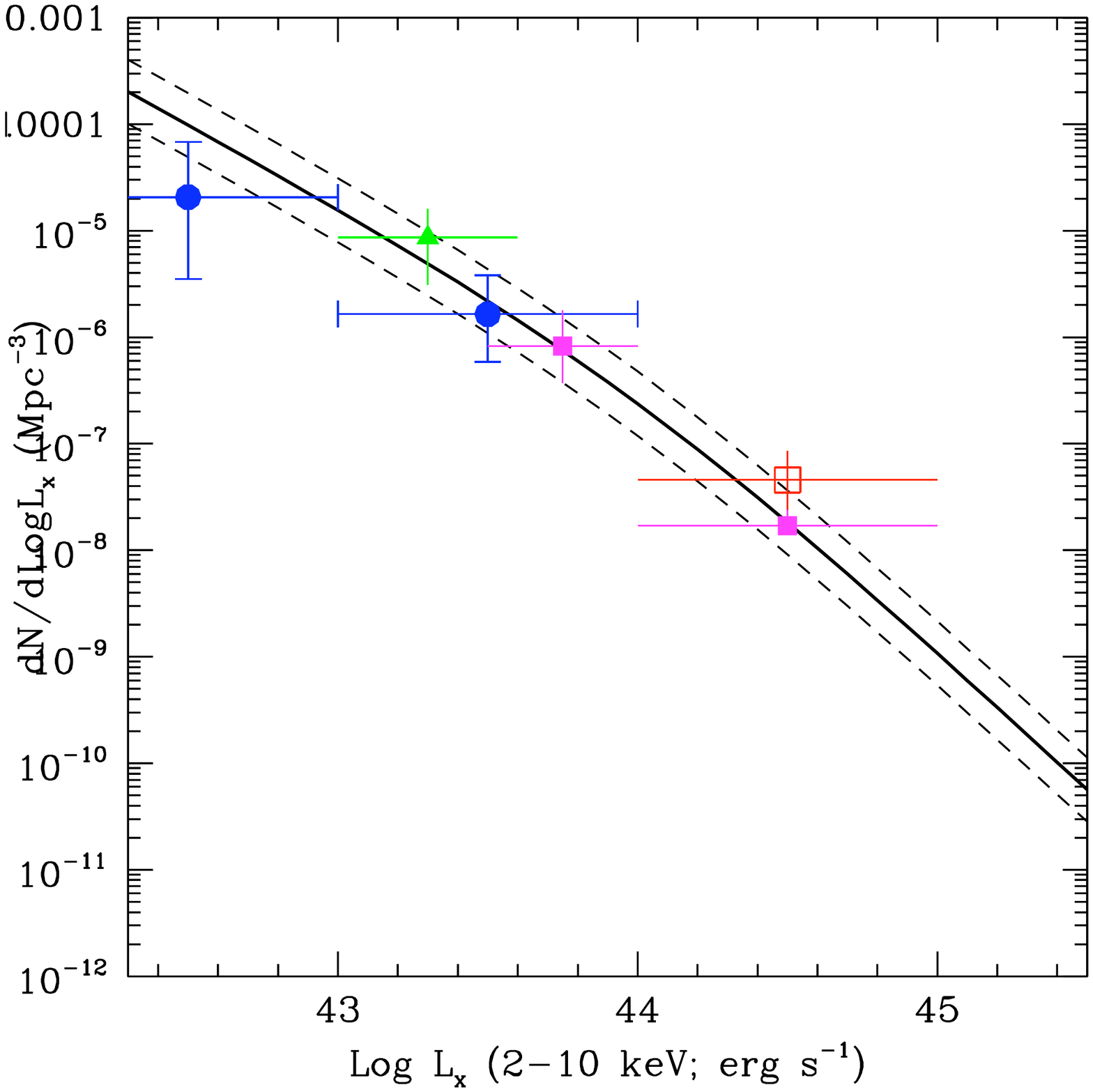} 
   \includegraphics[width=5.5cm]{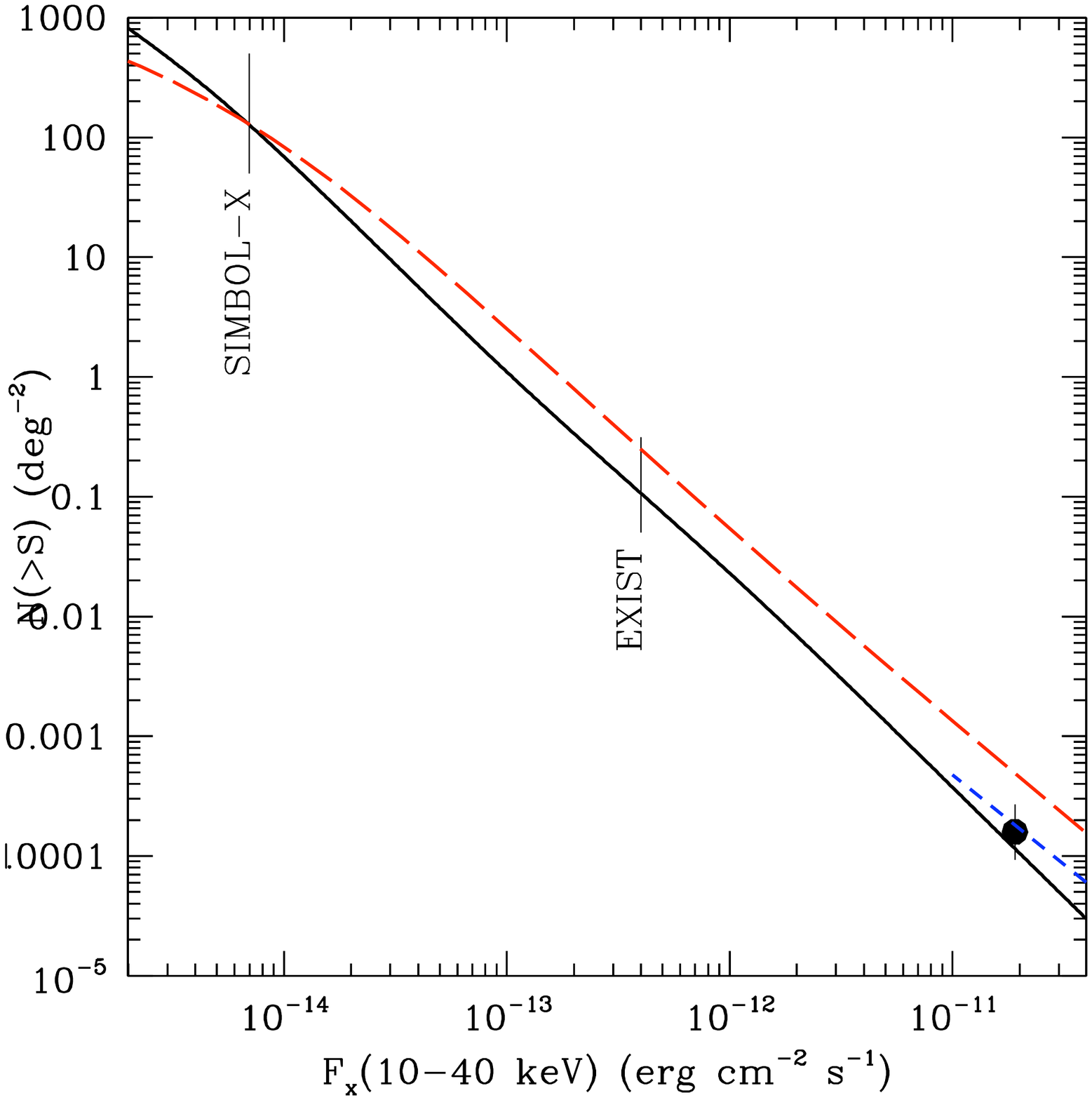}  
   \includegraphics[width=5.5cm]{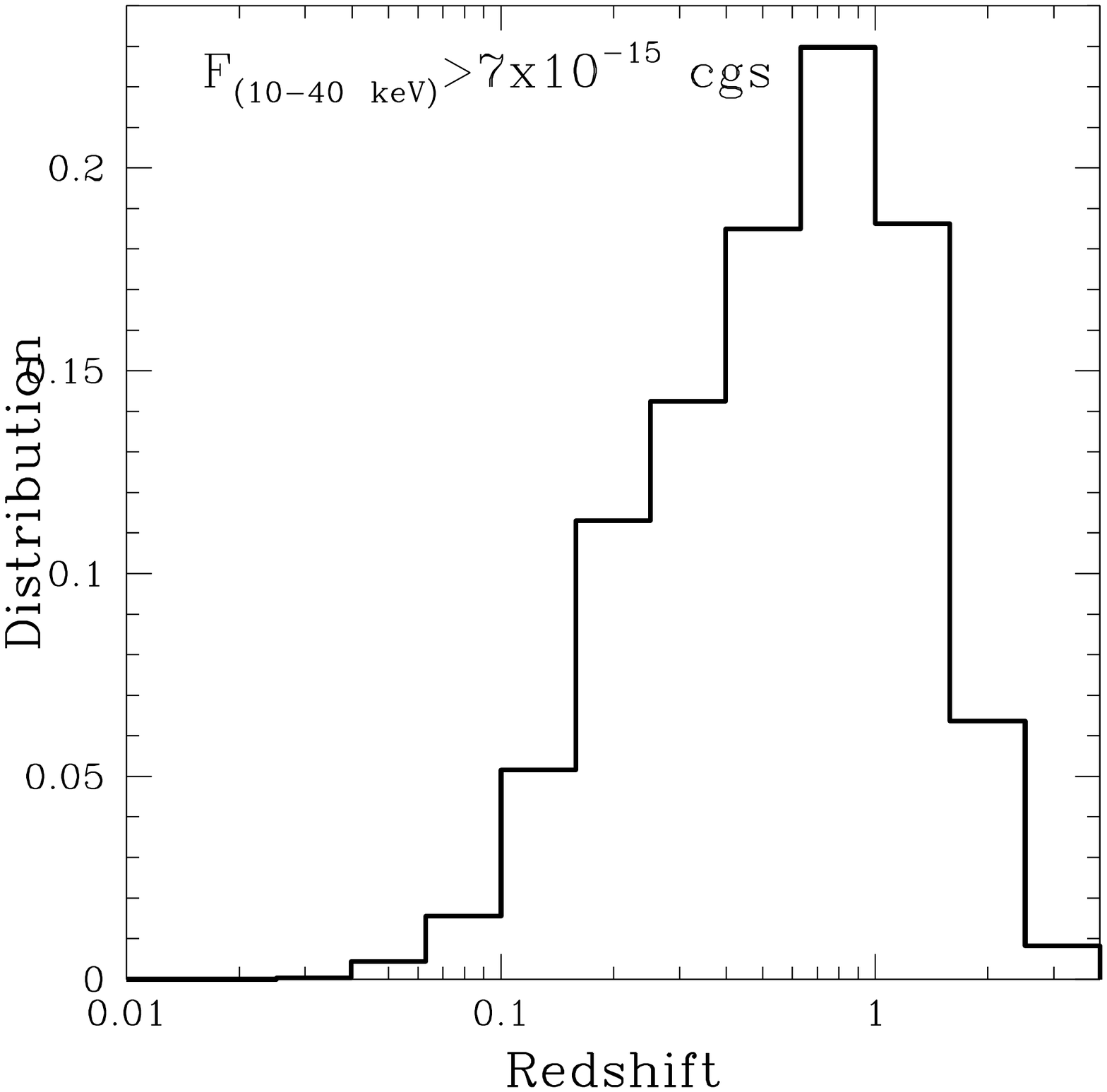}	 
\caption{{\it Left panel}:
The thick solid line represents the proposed intrinsic 2-10 keV XLF of Compton Thick AGN (at z=0), while  the two dashed lines correspond to an uncertainty of a factor two on the XLF normalization. The filled circles at
$L_{(2-10 keV)} \sim 3\times 10^{42}$ and  $\sim 3\times 10^{43}$ \es
represent  the density of local Compton Thick AGN derived using the INTEGRAL all sky survey discussed in \cite{sazonov2007}.  
The filled triangle represents the density of Compton Thick AGN with  $L_X \sim
(1-4)\times 10^{43}$ \es as derived by \cite{daddi2007}.
The filled squares represent the density of Compton Thick AGN  with  $L_X 
\sim (3-10)\times 10^{43}$ \es  and $L_X \sim (1-10)\times 10^{45}$ \es,  as
derived by  \cite{fiore2008}, while the open square at 
$L_X \sim (1-10)\times 10^{45}$ \es corresponds to the density of Compton Thick 
AGN as derived by \cite{alexander2008}. The densities of Thick AGN reported in \cite{daddi2007}, \cite{fiore2008} and \cite{alexander2008}
have been rescaled to z=0 using the best fit evolutionary model 
discussed in \cite{dellaceca2008}.
{\it Middle panel}: 
The solid line is the LogN($>$S)-LogS in the 10-40 keV energy range derived as explained in the text. 
We have also reported the expected flux limits of the All Sky
EXIST mission (``1 year survey data") and SIMBOL-X (1 Msec exposure).  
The filled dot at $\sim 2\times 10^{-11}$ \cgs represents the surface density of Compton Thick AGN obtained from the SWIFT/BAT survey, the short dashed line 
is the LogN($>$S)-LogS of Compton Thick AGN from \cite{treister2009} while 
the long-dashed line are the prediction from \cite{gilli2007} based on the XRB 
modeling.
{\it Right panel}: 
Expected redshift distribution for the Compton Thick AGN population at the SIMBOL-X flux limit of $F_{10-40 keV} \sim 7\times 10^{-15}$ \cgs
}
\end{figure}

\section{Predictions for the SIMBOL-X mission}

Having derived an estimate of the intrinsic XLF of Compton Thick AGN (at z=0) we can now make predictions for the Compton Thick AGN population that will be detected above 10 kev (in particular in the 10-40 keV energy range) with 
SIMBOL-X. 

By integrating the XLF reported in figure 1 ($L_X$ from $10^{42}$ to $10^{48}$ 
${\rm erg~\rm s}^{-1}$; 
z from 0 to 3) using the best fit cosmological evolution parameters reported in \cite{dellaceca2008} and  convolving the results with 
{\it a)} the modeling of X-ray spectra transmitted through 
Compton-thick absorbers as discussed in \cite {matt1999} and 
{\it b)} a flat $N_H$ distribution between $10^{24}$ and $10^{25}$ cm$^{-2}$, 
we predict the LogN($>$S)-LogS reported as solid line in figure 1 (middle panel). 

In this figure we also shown: 

{\it a)} the surface density of Compton Thick AGN obtained from the SWIFT/BAT survey (\cite{tueller2008})
\footnote{The SWIFT/BAT results are fully consistent with those obtained from the INTEGRAL surveys (L.Bassani, private communication)}. 
To this purpose we have used the 5 Compton Thick AGN (as revised in \cite{treister2009}, i.e. MKN3, NGC3281, NGC4945, NGC5728 and NGC7582) having $N_H$ between $10^{24}$ and $10^{25}$ \cm and $F_{14-195 keV} > 6\times 10^{-11}$ \cgs (where we have a flat 
sky coverage and a covered area of $\sim$ 30526 deg$^{-2}$). 
To convert the fluxes from the different energy bands we have used an absorbed
power law spectral model having photon index $\Gamma=1.9$ and $N_H\sim 3\times 10^{24}$ \cm ($F_{10-40 keV} \sim 0.32\times F_{14-195 keV}$);

{\it b)} the LogN($>$S)-LogS of Compton Thick AGN ($N_H$ between $10^{24}$ and $10^{25}$ cm$^{-2}$) as derived by \cite{treister2009} (short-dashed line) 
using SWIFT/BAT and INTEGRAL data and 

{\it c)} the predictions based on the XRB modeling (long-dashed line. R. Gilli, 
private communication). 

As can be clearly seen we find a very good agreement between our predictions and the measurements performed so far with SWIFT/BAT, making us confident that our predictions for the SIMBOL-X mission are robust. We also find that the 
predictions based on the XRB modeling are about a factor 3 above the observed value at $F_{10-40 keV} \sim 2\times 10^{-11}$ 
${\rm erg~cm}^{-2}~{\rm s}^{-1}$, probably requiring a revision on the current XRB model's parameters (see also \cite{treister2009}).

At the SIMBOL-X flux limit of $F_{10-40 keV}\sim 7\times 10^{-15}$ \cgs 
(1 Msec exposure) we predict a Compton Thick AGN surface density of 125 deg$^{-2}$, i.e. around 5 Compton Thick AGN for each SIMBOL-X deep field. 
Thus with about 10 deep fields we will be able to constrain the density 
of Compton Thick AGN with an error of about 15-20\% in the redshift range 
between 0.8 and 1.5, where the peak of the redshift distribution is expected 
(see Figure 1, right panel: z distribution).
For a typical SIMBOL-X exposure of 30 Ksec the flux limit (10-40 keV band) is expected to be $\sim 6\times 10^{-14}$ 
${\rm erg~cm}^{-2}~{\rm s}^{-1}$. 
At this flux limit heavily 
obscured AGN can be assembled by looking for serendipitous sources in pointed observations; considering a total number of $\sim 1000$ SIMBOL-X pointings we should collect ``for free" other $\sim 150$ serendipitous Compton Thick AGN spanning a redshift range from 0.05 to 0.6.

Finally it is worth noting that in the bright flux regime
(few times above $\sim 10^{-13}$
${\rm erg~cm}^{-2}~{\rm s}^{-1}$), a complementary mission to SIMBOL-X is represented by the Energetic X-ray Imaging Survey Telescope (EXIST; \cite {grindlay2005}) that is expected to perform an all sky survey.
At the "1 year survey flux limit" (5$\sigma$) of $F_{10-40 keV} \sim 4\times 10^{-13}$ \cgs the predicted surface densities of Compton Thick AGN is 
$\sim 0.1$ deg$^{-2}$ (see Figure 1, middle panel), so $\sim$ 3000 Compton Thick AGN should be detected at high galactic latitude (the large majority of them expected at $z<0.2$).
In this respect SIMBOL-X and EXIST are synergic and not competing for what concerns the cosmological studies of the Compton Thick AGN population: with EXIST it will be possible to derive the census (e.g. XLF) of these objects in the local Universe while the much deeper surveys that will be carried out with SIMBOL-X will be a fundamental and unique tool to study the evolution of their XLF up to cosmological distances ($z\sim 2$).   
As already stressed by \cite{treister2009} the study of these surveys at E$>10$ keV is {\bf the only way} to constrain, on firm and solid statistical bases, the census of heavily obscured accretion in AGN. 


\begin{theacknowledgments}
 
We acknowledge financial support from the MIUR (grant PRIN-MIUR 2006-02-5203)
and from the Italian Space Agency (grant n. I/088/06/0 and COFIS contract).

\end{theacknowledgments}


\bibliographystyle{aipproc}   



\end{document}